# 'Einselection' of Pointer Observables: The New H-Theorem?

Ruth E. Kastner[†]

16 June 2014

to appear in *Studies in History and Philosophy of Modern Physics*

ABSTRACT. In attempting to derive irreversible macroscopic thermodynamics from reversible microscopic dynamics, Boltzmann inadvertently smuggled in a premise that assumed the very irreversibility he was trying to prove: 'molecular chaos.' The program of 'Einselection' (environmentally induced superselection) within Everettian approaches faces a similar 'Loschmidt's Paradox': the universe, according to the Everettian picture, is a closed system obeying only unitary dynamics, and it therefore contains no distinguishable environmental subsystems with the necessary 'phase randomness' to effect einselection of a pointer observable. The theoretically unjustified assumption of distinguishable environmental subsystems is the hidden premise that makes the derivation of einselection circular. In effect, it presupposes the 'emergent' structures from the beginning. Thus the problem of basis ambiguity remains unsolved in Everettian interpretations.

1. Introduction

The decoherence program was pioneered by Zurek, Joos, and Zeh,[1] and has become a widely accepted approach for accounting for the appearance of stable macroscopic objects. In particular, decoherence has been used as the basis of arguments that sensible, macroscopic 'pointer observables' of measuring apparatuses are 'einselected' – selected preferentially by the environment. An important application of 'einselection' is in Everettian or 'Many Worlds' Interpretations (MWI), a class of interpretations that view all 'branches' of the quantum state as equally real. MWI abandons the idea of non-unitary collapse and allows only the unitary Schrödinger evolution of the quantum state. It has long been known that MWI suffers from a 'basis arbitrariness' problem; i.e., it has no way of explaining why the 'splitting' of worlds or of macroscopic objects such as apparatuses and observes occurs with respect to realistic bases that one would expect. That is, MWI does not explain why Schrodinger's Cat is to be viewed as 'alive' in one world and 'dead' in another, as opposed to 'alive + dead' in one world and 'alive – dead' in the other. Einselection, a shortened form for the phrase "environmentally induced superselection," is purported to come to the rescue by naturally steering such systems towards sensible 'pointer observables,' in which macroscopic objects and observers have determinate values with respect to the properties with which we are empirically familiar.

---
[†] UMCP Foundations of Physics Group; rkastner@umd.edu
[1] See References for key publications by Joos, Zeh, Zurek and co-authors.

This paper does not pretend to provide a comprehensive or exhaustive review of the decoherence program; many informative studies can be found in the literature (some key examples are in the listed in the References). Here I simply review the basic approach to 'deriving' einselection via decoherence, and point to a key step in the derivation that makes it a circular one. More sophisticated examples and arguments still depend on this assumption in one form or another, so it is sufficient for present purposes to deal with a simple example.

There are alternative decoherence approaches, in which it is noted that the 'pointer basis' depends on the choice of partition of all degrees of freedom into system + environment. Examples of this alternative approach can be found in Zanardi (2001) and Dugić and Jeknic-Dugić (2012). These approaches can be considered as a weaker version of 'einselection' in that they derive a pointer basis only relative to such a partition. However, this alternative form is strongly observer-dependent.[2] This paper is primarily aimed at refuting claims that the emergence of classicality proceeds in an observer-independent manner in a unitary-only dynamics; this is termed "Quantum Darwinism." Quantum Darwinism holds that the emergence of classicality is not dependent on any inputs from observers; it is the classical experiences of those observers that the decoherence program seeks to explain from first principles (e.g. Zurek, Riedel and Zwolak, 2013).

In addition, I will briefly discuss why the experiments purporting to demonstrate decoherence (as allegedly arising solely from unitary dynamics) do not actually do this. I also note that some form of irreversibility, such as nonunitary collapse, is necessary in order to remove the circularity.

2. Review of 'einselection' arguments

---

[2] In particular, Zanardi argues that there is no unique quantum-theoretic description of entanglement, and that whether or not a system can be described as entangled depends on how it is observed. This point can be sustained in the non-relativistic theory, in which quantum systems are primitive notions. However, it becomes harder to sustain when relativistic field theories are brought into play. An entangled two-photon Fock state is physically distinct from a two-photon Fock product state in terms of the basic field definition, since a Fock state specifies the energies of the photons. Also, it is well known that position is not a well-defined observable in QFT. Thus, the field picture breaks the basis arbitrariness, at least to some degree. An interpretational approach involving collapse in the field picture is explored in a separate work.

To review the alleged derivation of decoherence, I follow the straightforward presentation of Bub (1997). Consider the simple model presented in Zurek (1982), of a 2-level system S coupled to an environment E characterized by $n$ 2-level systems. The interaction Hamiltonian is

$$H_{int} = -\sum_j g_j R \otimes R_j \otimes \prod_{j' \neq j} I_{j'} \tag{1}$$

where $R$ is a system observable and $R_j$ are the observables of each of the $n$ environmental subsystems which couple to $R$ via the coupling constants $g_j$. The eigenstates of $R$ and $R_j$ are $|\pm\rangle$ and $|\pm\rangle_j$ respectively. If the initial state of S + E is

$$|\Psi(0)\rangle = (a|+\rangle + b|-\rangle) \otimes \prod_{j=1,n}(\alpha_j|+\rangle_j + \beta_j|-\rangle) \tag{2}$$

then the state at time $t$ is

$$|\Psi(t)\rangle = a|+\rangle \otimes \prod_{j=1,n}(\alpha_j e^{ig_j t}|+\rangle_j + \beta_j e^{-ig_j t}|-\rangle) + b|-\rangle \otimes \prod_{j=1,n}(\alpha_j e^{-ig_j t}|+\rangle_j + \beta_j e^{ig_j t}|-\rangle) \tag{3}$$

One then obtains the reduced mixed state $W_S$ of S by tracing over the environmental degrees of freedom in the density matrix $|\Psi(t)\rangle\langle\Psi(t)|$, to obtain

$$W_S = |a|^2|+\rangle\langle+| + |b|^2|-\rangle\langle-| + z(t)ab^*|+\rangle\langle-| + z^*(t)a^*b|-\rangle\langle+| \tag{4}$$

where

$$z(t) = \prod_{j=1,n}[\cos 2g_j t + i(|\alpha|^2 - |\beta|^2)\sin 2g_j t] \tag{5}$$

The goal of decoherence is to obtain vanishing of the off-diagonal terms, which corresponds to the vanishing of interference and the selection of the observable $R$ as the one with respect to which the universe purportedly 'splits' in an Everettian account. As observed by Bub, since the resulting mixed state is an improper one, it does not license the interpretation of the system's density matrix as representing the determinacy of outcome perceived by observers -- but that is a separate issue. The aim of this paper is to point out that the vanishing of the off-diagonal terms is crucially dependent on an assumption that makes the derivation circular.

As is apparent from (5), the off-diagonal terms are periodic functions that oscillate in value as a function of time. However, as Bub notes, $z(t)$ will have a very small absolute value, providing for very fast vanishing of the off-diagonal elements and a very long recurrence time for recoherence when $n$ is large, based on the assumption that *the initial states of the n environmental subsystems and their associated coupling constants are random*. But the randomness appealed to here is not licensed by the Everettian program, which states that the quantum state of the universe is that of a closed system that evolves only unitarily. The 'randomness' of the environmental systems does not arise from within the Everettian picture. When one forecloses that assumption, the decoherence argument fails – and with it, 'einselection,' which depends on essentially the same argument to obtain a preferred macroscopic observable for 'pointers.'

Of course, decoherentists argue that no observed system is ever 'closed' – meaning that it is interacting with its environment. However, the difficulty is that the 'openness' of the system is not actually available in the Everettian, unitary-only picture. The latter can only make an arbitrary division into system + pointer + environment (S+P+E). The total system S+P+E is closed, and within the Everettian picture this total system must therefore be described by the quantum state of the whole universe. The division S+P+E might be said to be non-arbitrary based on the *observed* fact that the environment appears to be made up largely of uncorrelated systems, but that crucially begs the question: MWI must be able to support this 'natural' division from within the unitary evolution only. After all, the whole point of the 'einselection' program is to demonstrate that the *observed* divisions arise naturally from within the theory. To assume the divisions we *already* see in the world and then demonstrate that, based on those assumed divisions, the divisions arise 'naturally,' is clearly circular.

The crucial point that does not yet seem to have been fully appreciated is this: in the Everettian picture, everything is always coherently entangled, so pure states must be viewed as a fiction -- *but that means that it is also fiction that the putative 'environmental systems' are all randomly phased*. In helping themselves to this phase randomness, Everettian decoherentists have effectively assumed what they are trying to prove: macroscopic classicality only 'emerges' in this picture because a classical, non-quantum-correlated environment was illegitimately put in by hand from the beginning. Without that unjustified presupposition, there would be no vanishing of the off-diagonal terms and therefore no apparent diagonalization of the system's reduced density matrix that could support even an approximate, 'FAPP' mixed state interpretation.

Thus, we see that the 'randomness of initial states' is logically equivalent to the assumption of 'molecular chaos' that permitted Boltzmann to 'prove' that entropy must always increase. Everettians invoking decoherence arguments to resolve their basis degeneracy argument face a similar 'Loschmidt's paradox': they are trying to derive a determinate macroscopic world of appearance, clearly divided along lines of preferred macroscopic observables, from a quantum dynamics that has no such preference for macroscopic observables.

Furthermore, the einselection program fares worse than H-theorem program for the following reason. Note that, while the second law of thermodynamics can be saved by assuming suitable initial conditions (i.e. a low entropy initial universal state), einselection cannot, because the initial condition is assumed to be a single universal wave function. So no recourse to boundary conditions is available for getting the einselection program onto non-circular footing.

The general problem with circularity in connection with 'einselection' has not gone entirely unnoticed. For example, Schlosshauer has this to say against a form of the circularity charge:

> "The clear merit of the approach of environment- induced superselection lies in the fact that the preferred basis is not chosen in an *ad hoc* manner simply to make our measurement records determinate or to match our experience of which physical quantities are usually perceived as determinate (for example, position). Instead the selection is motivated on physical, observer-free grounds, that is, through the system-environment interaction Hamiltonian. The vast space of possible quantum-mechanical superpositions is reduced so much because the laws governing physical interactions depend only on a few physical quantities (position, momentum, charge, and the like), and the fact that precisely these are the properties that appear determinate to us is explained by the dependence of the preferred basis on the form of the interaction. The appearance of "classicality" is therefore grounded in the structure of the physical laws—certainly a highly satisfying and reasonable approach…" (Schlosshauer 2004)

But this response fails to acknowledge or address the true source and extent of the circularity problem facing 'einselection.' Yes, the account could be considered observer-free based on the apparent availability of a well-defined physical interaction Hamiltonian between the system and its environment. But the problem is not so much a lack of observer-independence as it is a *failure to account for the initial independence of the environment from the system that it is measuring*. That is, even if it is true that the system's only correlation with the environment is via the interaction Hamiltonian, and the environmental systems are randomly phased with respect to each other, these conditions cannot be explained from within the Everettian account: in that account, random phases are fictions. And these conditions are crucial to 'deriving' decoherence and the appearance of classicality in the no-collapse, unitary-only Everettian picture. Thus, in that picture, apparently *de facto* classicality is crucial to deriving classicality. This should not be viewed as satisfying, any more than it can be viewed as satisfying to 'derive' the H-Theorem from a 'molecular chaos' assumption, even if there (*de facto*) really is molecular chaos.

It should be noted that Fields (2010) has also raised a similar charge of circularity against the program of 'quantum Darwinism' based on its division of the world into S+P+E based on macroscopic or classical preferences. However, his argument is focused on the impossibility of demonstrating coding redundancy in the environment, while the present focus is on the impossibility of deriving decoherence from an Everettian picture.

3. If the derivations are circular, why do experiments seem to 'confirm' decoherence?

There are many experiments reflecting the *de facto* phenomenon of decoherence. These experiments are often quite sophisticated and elegant. For example, Raimond, Brune, and Haroche (2001) obtain mesoscopic superpositions of coherent electromagnetic field states $|\alpha\rangle$ and show that the superposition decays in accordance with the predictions of decoherence theory as $\alpha$ (the square root of the average photon number in the field) is varied. However, the ability to show that interference can be lost, even in a controlled manner, does not corroborate the 'einselection' program, since that program is circular and therefore cannot get off the ground. The studied field system in the RBH experiment is presumed to be already decohered from its environment. So this again begs the question: if the relevant phases are indeed randomized, how did they get randomized in the first place? The answer cannot be found in the unitary-only Everettian program.

The basic point here is that any putative experimental corroboration of Quantum Darwinism is illusory, just as an experimental corroboration of the H-theorem 'proof' would be illusory. Yes, we observe that entropy increases, but that is an empirical fact, not one that is derivable via reversible physical laws (without special boundary conditions). Yes, we observe decoherence, but that is also an empirical fact not derivable from a unitary-only dynamics. Both the H-theorem and unitary-only derivations of 'einselecton' are crucially dependent on assuming, in effect, what they are trying to prove.

On the other hand, if quantum systems actually do undergo irreversible physical collapse, then their phases could naturally become randomized via this nonunitary process. Thus, collapse could be the missing ingredient in both the H-Theorem derivation and in einselection derivations. In order to remove the circularity in einselection derivations via collapse, one would need to provide a non-arbitrary account of entanglement and an observer-independent definition of the physical systems to which the quantum description is being applied. As suggested in footnote 2, working in the relativistic field picture may be a promising way to achieve these goals.

4. Conclusion

It is often claimed that unitary-only dynamics, together with decoherence arguments, can explain the 'appearance' of wave function collapse, i.e, that Schrodinger's Cat is either alive or is dead. This however is based on implicitly assuming that macroscopic systems (like Schrodinger's Cat himself) are effectively already 'decohered,' since the presumed phase randomness of already-decohered systems is a crucial ingredient in the 'derivation' of decoherence. Thus decoherence arguments alone do not succeed in providing for the emergence of a classical world, nor for the necessary preferred basis of splitting in an Everettian account, and their explanatory benefit is illusory.

Acknowledgements.
The author would like to thank an anonymous referee for valuable comments.